\documentclass[conference,compsoc]{IEEEtran}
\usepackage{floatrow}
\usepackage{hyperref}
\usepackage{listing}
\usepackage{opera}
\usepackage[sort,compress,numbers]{natbib}
\usepackage[letterpaper]{geometry}
\usepackage{tcolorbox}
\usepackage{fullwidth}
\usepackage{endnotes}
\usepackage[printwatermark]{xwatermark}

\let\footnote=\endnote

\newcommand{\contentbox}[2]{
\begin{figure*}
\begin{tcolorbox}[colback=Thistle!5!white,colframe=Thistle!75!black,fonttitle=\bfseries,title={#1}]
	#2
\end{tcolorbox}
\end{figure*}
}

\definecolor{codegreen}{rgb}{0,0.6,0}
\definecolor{codegray}{rgb}{0.5,0.5,0.5}
\definecolor{codepurple}{rgb}{0.58,0,0.82}
\definecolor{backcolour}{rgb}{0.95,0.95,0.92}

 \lstdefinestyle{mystyle}{
	backgroundcolor=\color{backcolour},
	commentstyle=\color{codegreen},
	keywordstyle=\color{magenta},
	numberstyle=\tiny\color{codegray},
	stringstyle=\color{codepurple},
	basicstyle=\footnotesize\ttfamily,
	breakatwhitespace=false,
	breaklines=true,
	captionpos=b,
	keepspaces=true,
	numbers=left,
	numbersep=5pt,
	showspaces=false,
	showstringspaces=false,
	showtabs=false,
	tabsize=2
}

\renewcommand{\baselinestretch}{0.98}
\usepackage[printwatermark]{xwatermark}

\begin{document}
\title{Sharing and Preserving Computational Analyses for Posterity with \texttt{encapsulator}}

\author{\IEEEauthorblockN{Thomas Pasquier\IEEEauthorrefmark{1}\IEEEauthorrefmark{2},
		Matthew K. Lau\IEEEauthorrefmark{3},
		Xueyuan Han\IEEEauthorrefmark{2},
		Elizabeth Fong\IEEEauthorrefmark{4},\\
		Barbara S. Lerner\IEEEauthorrefmark{4},
		Emery Boose\IEEEauthorrefmark{3},
		Merc{\`e} Crosas\IEEEauthorrefmark{5},
		Aaron Ellison\IEEEauthorrefmark{3}, 
		Margo Seltzer\IEEEauthorrefmark{2}
	}
	\IEEEauthorblockA{\IEEEauthorrefmark{1}Department of Computer Science and Technology\\
		University of Cambridge, Cambridge, UK\\}
	\IEEEauthorblockA{\IEEEauthorrefmark{2}School of Engineering and Applied Sciences\\
		Harvard University, Cambridge, USA\\}
	\IEEEauthorblockA{\IEEEauthorrefmark{3}Harvard Forest,\\
		Harvard University, Petersham, USA}
	\IEEEauthorblockA{\IEEEauthorrefmark{4}Department of Computer Science,\\
		Mount Holyoke College, South Hadley, USA}
	\IEEEauthorblockA{\IEEEauthorrefmark{5} Institute for Quantitative Social Science,\\
		Harvard University, Cambridge, USA}
}

\maketitle

\begin{abstract}
Open data and open-source software may be part of the solution to science’s “reproducibility crisis,” but they are insufficient to guarantee reproducibility.
Requiring minimal end-user expertise, \texttt{encapsulator} creates a \emph{``time capsule''} with reproducible code in a self-contained computational environment.
\texttt{encapsulator} provides end-users with a fully-featured desktop environment for reproducible research.
 \end{abstract}

\IEEEpeerreviewmaketitle

\section{Introduction}
\label{sec:introduction}
Reproducibility has become a recurring topic of discussion in many scientific disciplines~\cite{baker20161}. Although it may be expected that some studies will be difficult to reproduce, recent conversations highlight important aspects of the scientific endeavor that could be improved to facilitate reproducibility. 
Open data and open-source software are two important parts of a con-certed effort to achieve reproducibility~\cite{gezelter2015open}. However, multiple publications point out the short-comings of these approaches~\cite{garijo2013quantifying, joppa2013troubling}, such as the identification of dependencies, poor documentation of the installation processes, “code rot,” failure to capture dynamic inputs, and technical barriers.

\begin{figure*}[!htbp]
	\centering
	\begin{floatrow}
		\ffigbox[0.5\FBwidth]
		{\centering\includegraphics[width=\columnwidth]{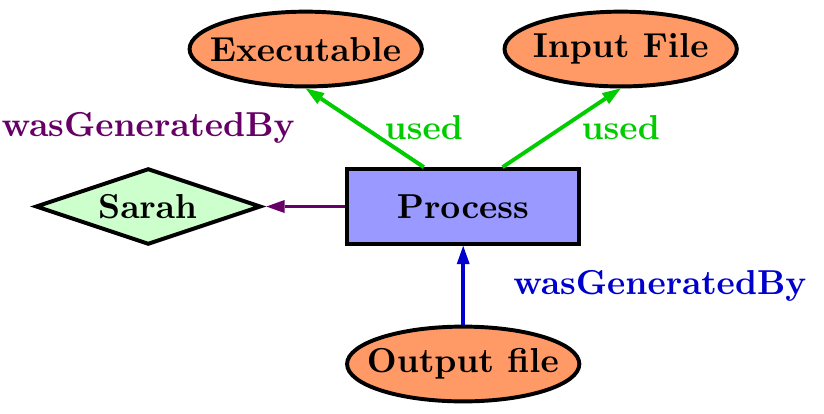}}
		{\caption{A simple W3C PROV-DM compliant provenance graph.}
			\label{img:provenance:w3c}}
		
		\ffigbox[0.5\FBwidth]
		{\includegraphics[width=\columnwidth]{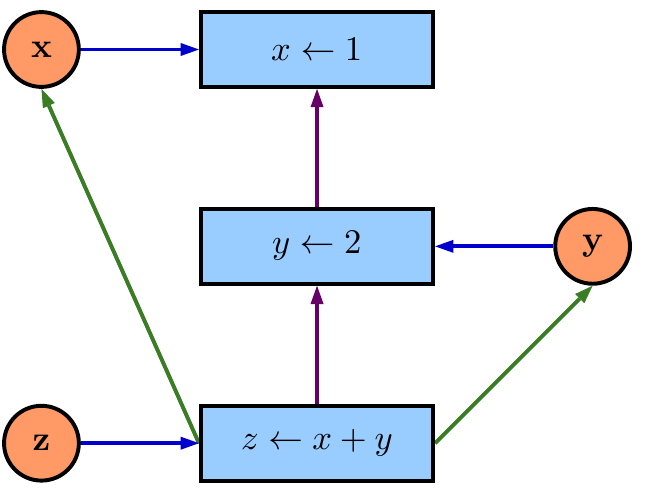}}
		{\caption{A simple provenance graph for an R script.}
			\label{img:provenance}}
	\end{floatrow}
\end{figure*}

In prior work~\cite{pasquier2017}, we pointed out that open data and open-source software alone are insufficient to ensure reproducibility, as they do not capture information about the computational execution, i.e., the ``process'' and context that produced the results using the data and code.
In keeping with the ``open'' culture, we defined open-process as the practice of both sharing the source and the input data and providing a description of the entire computational environment, including the software, libraries, and operating system used for an analysis.
We suggested the use of data provenance~\cite{carata2014primer}, formalized metadata representing the execution of a computational task and its context (\eg dependencies, specific data versions, and random or pseudo-random values), which can be captured during computation.

We view data provenance as key to addressing these issues, yet still insufficient.
We need tools that leverage provenance to put capabilities, not complex metadata, into scientists’ hands. 
We build on recent developments that address this need, such as executable papers~\cite{strijkers2011toward} and experiment packaging systems, \eg ReproZip~\cite{chirigati2016reprozip}. 
We propose a solution for scientists running small-to-medium-scale computational experiments or analyses on commodity machines.
Although tools exist to cover analyses done using spreadsheet programs (further dis-cussed in the Challenges section), we intentionally do not cover that space, as it has inherent barriers to transparency and identification of the source of errors~\cite{cunha2012towards, ziemann2016gene}.
Similarly, we do not attempt to address the reproducibility of large-scale computational analysis.

We present a \emph{time capsule} for small-to-medium-scale computational analysis. 
This \emph{time capsule} is a self-contained environment that allows other scientists to explore the results of a published paper, reproduce them, or build upon them with minimal effort. 
We automatically curate the scientist's code to extract only those elements pertinent to a particular figure, table, or dataset.

\section{Data Provenance}
\label{sec:provenance}
Data provenance~\cite{carata2014primer} has the potential to address some of the challenges related to reproducibility. 
Indeed, to assess the validity or quality of information, it is necessary to understand the context of its creation. Unfortunately, digital artifacts frequently omit or hide much of the context in which they were created. 
As an example, many of us have been guilty of sharing code we developed on our machines that our colleagues could not run, because we often work in the same environment for months or years, forgetting about software and libraries we have installed over time.

Meanwhile, small differences in a computational pipeline can lead to vastly different results.
For example, different analyses of the same dataset of carbon flux in an Amazonian forest ecosystem differed in their estimates by up to 140\%~\cite{Ellison2006}, amounting to differences of up to 7 tons of carbon in an area of the size of a football field. 
This example highlights the significant impact of small differences in code, especially when analyses or models contain user-defined or interactive (\eg multiplicative) terms. 
Seemingly small changes to inputs or in the computational pipeline can lead to large differences in results, impeding their reproducibility and verification.

Data provenance is a formal representation of the context and execution of a computation. This information is represented as a directed acyclic graph (DAG), a structure amenable to computational analysis.
We use the World Wide Web Consortium (W3C) standard for data provenance: PROV-DM. \autoref{img:provenance:w3c} shows a simple provenance graph. 
Vertices represent enti-ties (representing data), activities (representing actions or transformations), and agents (repre-senting users or organizations). 
In \autoref{img:provenance:w3c}, a process, controlled by Scientist Sarah, uses an executable function (\ie a program) and an input file (\ie data) to generate an output.

Provenance can be captured at various levels of a system, such as in libraries explicitly called by a program, in a language interpreter, in system libraries, or in the operating system. 
The specific capture approach produces subtly different types of provenance: observed provenance is deduced by a system that monitors execution, whereas disclosed provenance is created explicitly by software that understands the semantics of the computations performed~\cite{braun2006issues}.
\texttt{encapsulator} uses observed provenance capture, which reveals the inner workings of an analysis script by collecting fine-grained provenance.

When provenance is captured for a scripting or programming language, the provenance DAG represents relationships among inputs, outputs, transient data objects, and statements. For example, \autoref{img:provenance} illustrates a provenance graph of the following R script:
\begin{lstlisting}[language=R, style=mystyle]
x <- 1
y <- 2
z <- x + y
\end{lstlisting}
In the figure, the blue rectangles correspond to statements in the language; the orange circles correspond to data items (\ie inputs, outputs, or transient objects); the purple arrows show the control flow, representing the precise sequence of steps taken while executing the program; and the blue and green arrows show data dependencies (\ie the data used by an operation, and the data generated by an operation respectively).

The provenance DAG illustrates data dependencies (\ie what input generated a given output), software dependencies (\ie on what libraries a script depends), and information about the structure of a program. We next discuss how we use provenance DAGs to generate a \emph{time capsule}.
 
\section{Creating a Time-Capsule}
\label{sec:capsule}
\begin{figure}[t]
	\centering
	\includegraphics[width=0.8\columnwidth]{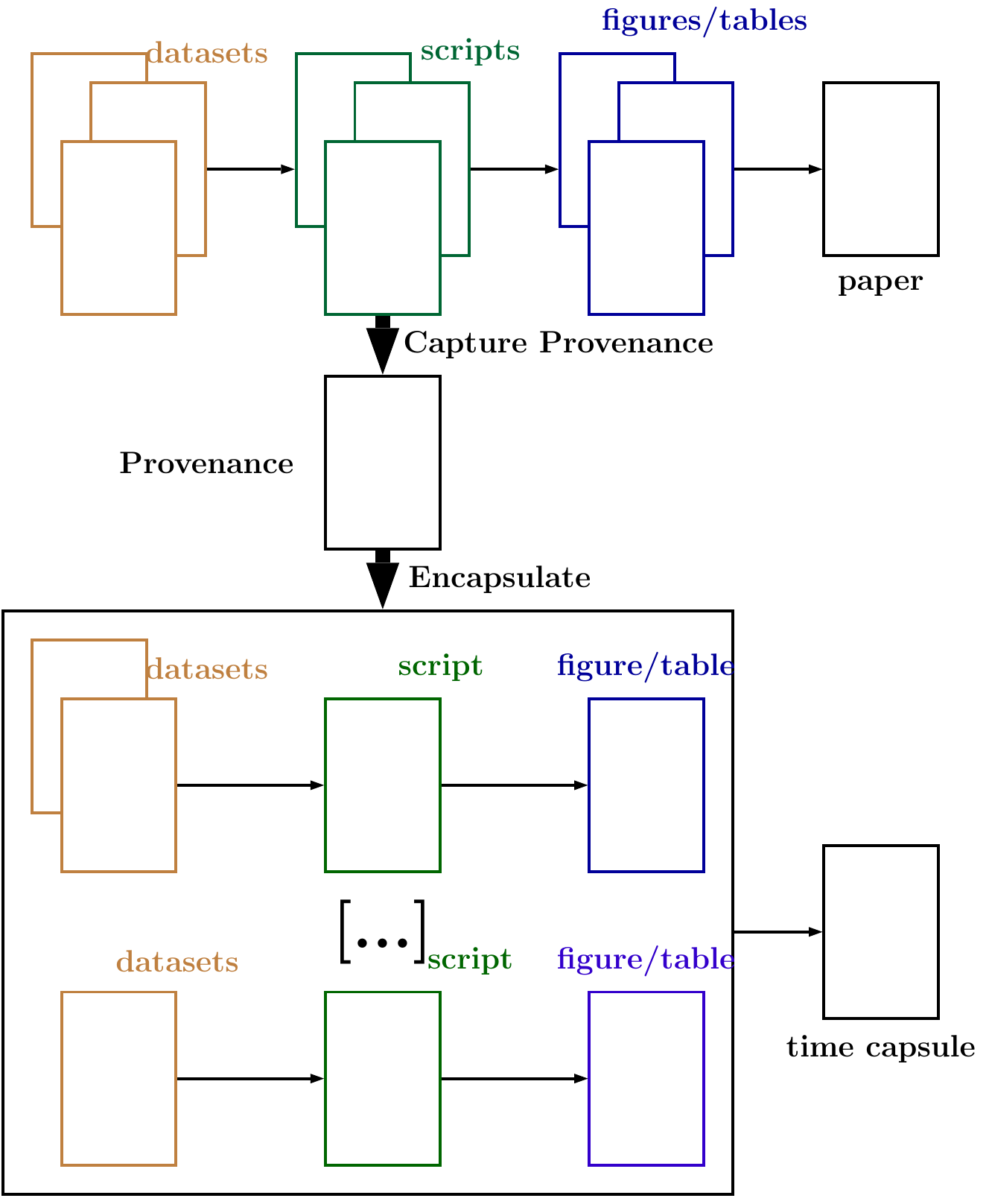}
	\caption{The encapsulation process.}
	\label{img:encapsulate}
\end{figure}

Provenance alone provides a ``picture'' of a computational context, yet we want to provide an active artifact that can reproduce a computational context: the time capsule. 
\autoref{img:encapsulate} illustrates the two phases involved in creating a time capsule from the provenance collected during execution: 
1) curate the script to identify the precise lines of code and input data needed to produce a result; 
2) build the time capsule containing the previously generated artifact and the environment necessary to reproduce it.

\subsection{Curating the code}

Science is, by its very nature, an iterative process. 
The task of cleaning and analyzing data is a stark example of this. 
The data obtained from scientific instruments or other measurements of the physical world are frequently a superset of the data a scientist wants to analyze. 
The first step in computation or analysis is often to ``process'' raw data to produce something that can be analyzed to answer a specific scientific question. 
This processing typically includes deciding how to handle missing data values, extracting parts of the data, computing new data from pieces of raw data, \etc 
A scientist typically performs many such operations, not all of which end up being useful.
Additionally, code evolves and accretes over time as scientists try different ways to interpret or analyze the data. 
False starts and abandoned analyses frequently persist in the final scripts that scientists use. 
The result is that code often contains a complex and evolving story of what transpired, rather than a clear, straight-line path from data to discovery. 
Although this history may be interesting, it may lead to confusing and difficult-to-understand code.

The first phase of \texttt{encapsulator} takes as input the provenance of the computation’s execution, including all the false starts and abandoned attempts, and produces a curated script corresponding to the generation of a specific result. 
Such a curated script contains the minimum sufficient code to generate the output. 
Therefore, to understand a specific result, one can examine the curated version, rather than having to wade through potentially large amounts of irrelevant code.

\begin{figure}[t]
	\centering
	\includegraphics[width=0.2\columnwidth]{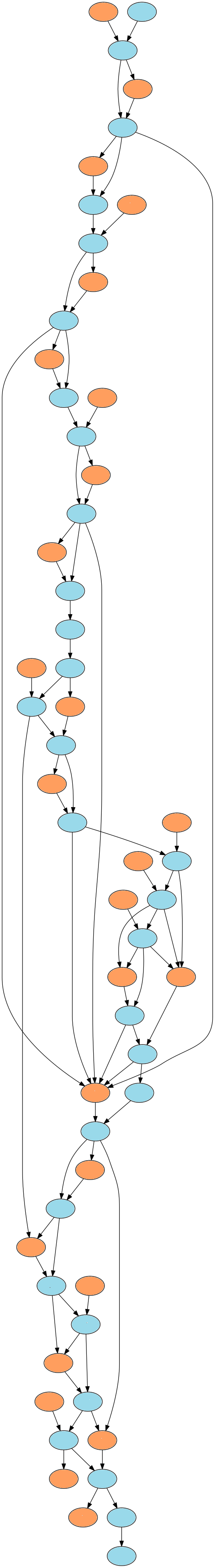}
	\caption{The provenance graph corresponding to a small R
		script (around 60 lines of code).}
	\label{img:beforepath}
\end{figure}

\begin{figure}[t]
	\centering \includegraphics[width=0.8\columnwidth]{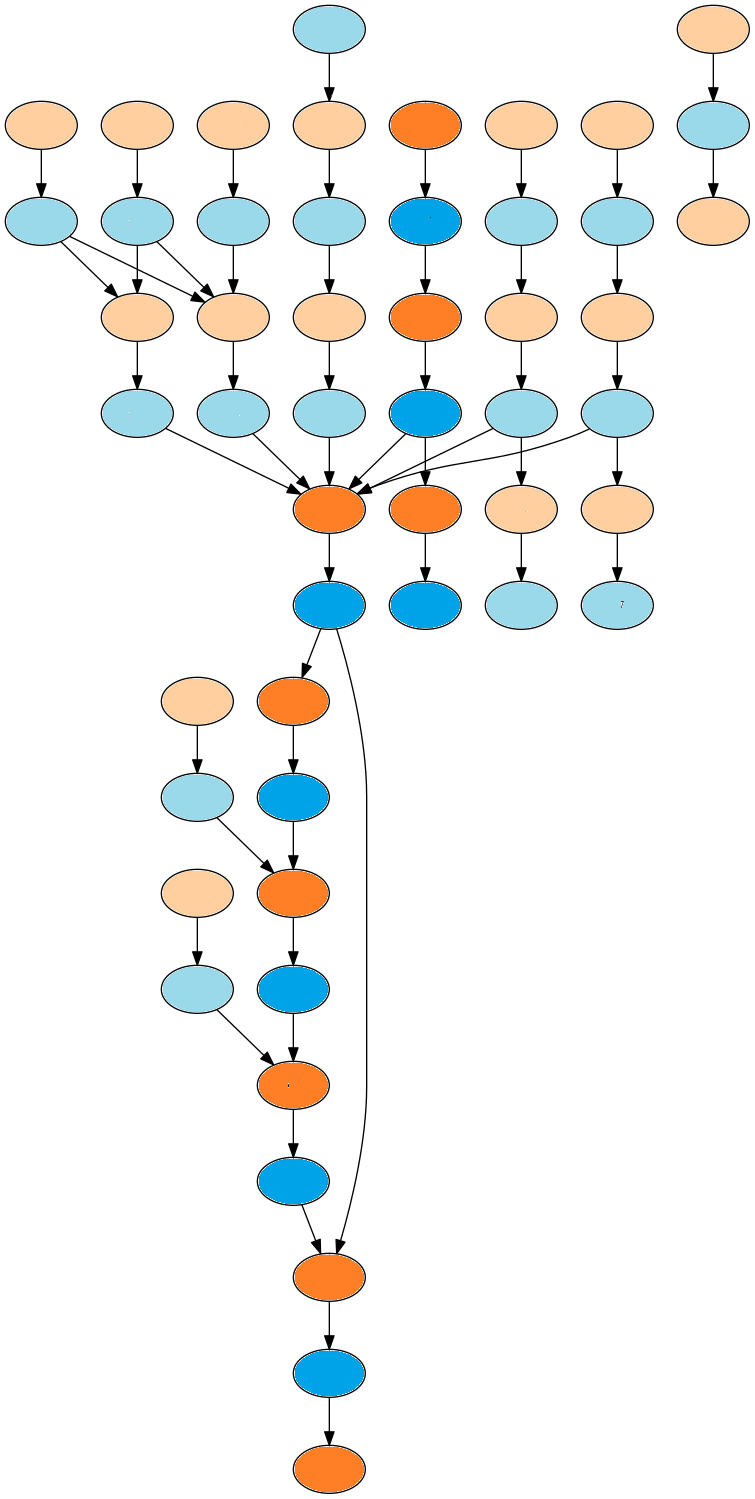}
	\caption{The data dependencies tranformation of the provenance graph shown in \autoref{img:beforepath}.}
	\label{img:path}
\end{figure}

To generate the minimal ``cleaned'' code, we analyze the provenance graph. 
Intuitively, the operations relevant to the generation of a figure or table are those connected in the DAG through data dependencies to the output. 
First, we trim the provenance graph by deleting control flow, considering only data dependencies. 
For example, the provenance graph illustrated in \autoref{img:beforepath} is transformed into a set of data dependency graphs shown in \autoref{img:path}. 

In a data dependency graph, orange nodes represent inputs, outputs, or transient data, and blue nodes represent operations on data items. 
As we examine data dependencies in \autoref{img:path}, we alternate between data items and operations. 
The code necessary to generate an output (at the top of the figure) is the ordered set of operations present on all paths starting from the output in the original source code (more intense colored nodes in \autoref{img:path} show an example of such a path for a given output). 
Similarly, the inputs necessary to generate an output are those encountered while traversing those paths. 
We generate the final, curated code by retaining all the operations on the paths in the graph leading to the output of interest, and then perform a final pass over the provenance DAG to identify all the required libraries. Once the final code has been generated, we run a source-code formatting tool (\texttt{formatR} for R scripts) to bring the code closer to best practices. 
We repeat these steps for every output of interest until we have generated a curated script for each. 
The inputs used to generate the selected outputs are identified and saved as part of the time capsule. 
We have made available (see \url{http://provtools.org/}) a standalone R library (Rclean \url{https://cran.r-project.org/web/packages/Rclean/}) implementing the mechanism described here.

\subsection{Building the time capsule}

Having shown how we produce curated scripts, we next explain how to construct a \emph{time capsule}, leveraging freely-available tools wherever possible. 
Our goal is to generate a self-contained environment that most scientists can use. 
This leads to the following requirements:
\begin{itemize}
	\item The environment should present a user interface familiar to scientists;
        \item Encapsulation and use (de-encapsulation) of time capsules must require minimal technical expertise;
        \item The installation process itself must also require minimum intervention and technical knowledge;
        \item Time capsules, their installation, and re-execution must be platform-independent.
\end{itemize}
We demonstrate through a practical scenario how well we meet those requirements in the next section.

Based on those criteria, we selected virtual machines (VMs) as the self-contained environment for our time capsules (\ie their behavior and content is independent of the guest machine, and will remain identical over time). 
As one of the main barriers to reproducibility is technical, we want to avoid introducing additional technical complexity. 
Software, such as \texttt{VirtualBox} (see \url{https://www.virtualbox.org/}), has made VMs an easy-to-use, ``push button'' technology, and it is possible to use a user-friendly interface to run a virtualized desktop with almost no technical knowledge. 
To most scientists, a VM will appear as a desktop environment similar to the one they use every day. 
To facilitate ease of adoption, we make sure that the time capsule contains all the tools scientists need to usefully interact with the computational process.

We use \texttt{Vagrant} (see \url{https://www.vagrantup.com/}) infrastructure and software to build, share, and distribute time capsules. 
Its VM provisioning is akin to that of \texttt{Docker} for containers. 
To provision a VM, one simply writes a script specifying the base VM (a pre-configured image), additional software, and files that should be installed. 
This is completely transparent to scientists: \texttt{encapsulator} generates a Vagrant file based on the information extracted from the provenance data in the previous phase. 
Although users can (optionally) customize the provision script, such customization should never be necessary. 
In the current prototype, the time capsule is Linux-based, as we leverage its package manager; other operating systems present licensing challenges (discussed in \autoref{sec:challenges}). 
However, the creation of the time capsule itself can be done from experiments running on Windows, Mac, or any Linux distributions.

The provenance capture is achieved through program introspection using ProvR (see \url{http://provtools.org/}). 
This presents some restrictions regarding the amount of system details that can be captured. 
In the current proof of concept implementation, we rely on the package manager of the Fedora Linux distribution (see \url{https://fedoraproject.org/wiki/dnf}) to install the system dependencies required by a specific version of an R library. 
We are exploring the possibility of complementing our provenance source using CamFlow~\cite{pasquier2017practical} (see \url{http://camflow.org/}) to capture system level provenance in the Linux operating system. 
However, it must be noted that system-level provenance capture in closed-source operating systems remains a challenge.

During encapsulation, the scripts created in the first phase run in the time-capsule environment. 
Their outputs are compared to those from the original script (\ie the one run on the host machine) to ensure that they are identical. 
Once the encapsulation is finished, the VM is packaged, ready to be shared. 
This VM contains individual R scripts for each selected figure, along with the datasets used as inputs. 
The current prototype relies on \texttt{Vagrant}’s cloud platform to host the VM.
 
\contentbox{Alternatives to \texttt{encapsulator}}{
Some systems are designed to reproduce complex workflows running on grid or cloud infrastructures (e.g., \texttt{Kepler}~\cite{altintas2004kepler}), and fill a related, but distinct niche. Indeed, \texttt{encapsulator} is intended to support research run on single commodity machines, which accounts for a significant proportion of research results in a number of fields. Systems designed for particular domains already exist (e.g., \texttt{GenePattern}~\cite{reich2006genepattern}, and \texttt{Galaxy}~\cite{giardine2005galaxy}), but the role of encapsulator is to provide a general approach.

\texttt{ReproZip}~\cite{chirigati2016reprozip} and \texttt{CDE}~\cite{howe2012cde} are directly comparable to \texttt{encapsulator}. However, they use system calls to identify dependencies and package experiments. Therefore, computations first must be run in Linux before they can be packaged. This may prove problematic for many scientists who do not use Linux. \texttt{encapsulator} relies on language-level observed provenance and is not subject to such limitations.

The main difference between \texttt{encapsulator} and alternative tools is the focus on ease of use. Modifying packaged computations generated by the alternatives may require a relatively high level of technical skill. \texttt{encapsulator} builds a fully functional, self-contained environment that is easy for scientists to navigate. The list presented here is succinct, but we maintain online a list of open-source provenance tools including some designed for reproducibility and replication purposes (see \url{https://projects.iq.harvard.edu/provenance-at-harvard/tools}). }

\section{Using Encapsulator}
\label{sec:example}
Consider the following scenario: Sarah is a young and brilliant scientist who would like to make her research results available to the community, allow reviewers to easily verify her results, and encourage others to build on them. 
Prof. O'Brien is a reviewer, interested in verifying Sarah's findings.
John is a scientist from a near future who wishes to use Sarah's results. 

The ``messycode'' examples (see \url{https://github.com/ProvTools/encapsulator}) illustrate several ``lazy coding practices'' that scientists, including Sarah, often use when writing code for models and analyses:
\begin{itemize}
\item near ``stream-of-consciousness'' coding that follows a train of thought in script de-velopment;
\item output to console that is not written to disk;
\item intermediate objects that are abandoned;
\item library and new data calls throughout the script;
\item output written to disk but not used in final documents;
\item code is not modularized;
\item code that is syntactically correct but not particularly comprehensible.
\end{itemize}

At this stage, we assume that Sarah has finished her computations, built the figures and tables for her paper, and has the paper ready for submission. 
She is aware of open-data repositories, such as \texttt{Dataverse} (see \url{https://dataverse.org/}), and source-code repositories, such as \texttt{GitHub} (see \url{https://github.com/}), but she knows they may not be sufficient to make her code truly re-usable. 
In the past, when she tried to re-use code written by other scientists, she often discovered that it was poorly documented and hard to use. 
She also constantly found herself baffled by questions such as what external packages the computation depends on, where to obtain those dependent files and libraries, and what parameters were used to obtain the published results. 
Trying to figure out these details resulted in her wasting countless hours. 
She would like to save other scientists from these challenges, so that they can more easily build upon her work.

Sarah wants a ``picture'' of the context of her computations that allows anyone to reproduce them. 
Provenance captured by tools such as \texttt{provR} (see \url{http://provtools.org/}) for R scripts contains the following information, represented as nodes or node attributes in a DAG:
\begin{itemize}
	\item inputs;
	\item outputs;
	\item transient data objects and their values;
	\item operations;
	\item library dependencies.
\end{itemize}
This information facilitates depiction of the development environment, accurately capturing, for example, random seeds used and the version of a library that was required by the system. 
Although this picture is important, it could prove difficult for John or Prof. O'Brien to use it to create an environment in which Sarah's computations can be reproduced. 
They may not have the required expertise or the required version of a library has become unavailable. 
Thus, Sarah wants her experiments to be preserved in a time capsule.

Sarah decides to use \texttt{encapsulator}. She needs to install it and its dependencies: \texttt{VirtualBox} and \texttt{Vagrant}. On her Mac laptop, she can do this:
\begin{lstlisting}[language=bash, style=mystyle, caption={Installing \texttt{encapsulator} and its dependencies.}, label=listing:install]
brew install ruby
gem install encapsulator
encapsulator --install mac
\end{lstlisting}
The next step is to examine her R script and determine what outputs she wants to include in her time capsule. She can find out what the possibilities are using \texttt{encapsulator}'s info capability:
\begin{lstlisting}[language=bash, style=mystyle, caption={Obtaining a summary of an R script.}, label=listing:info]
encapsulator --info sarah.R
\end{lstlisting}
This generates the following output:
\begin{lstlisting}[language=bash, style=mystyle, caption={Installing \texttt{encapsulator} and its dependencies.}, label=listing:infotxt]
Files
-----
Input july_biomass_survey.csv
Input dataset_v2_june_from_collaborator1.csv
Output save1.csv
Output fig1_biplot.png
Output fig1_biplot_v2.png
Output fig2_biplot.png

Packages
--------
base v3.4.0
gdata v2.18.0
lattice v0.20-35
permute v0.9-4
txtplot v1.0-3
vegan v2.4-3
\end{lstlisting}

Sarah included only \texttt{fig1\_biplot\_v2.png}
and \texttt{fig2\_biplot.png} in her article, so she wants to generate
a
\emph{time capsule} containing only the code (see supplementary material) needed to generate those two images:
\begin{lstlisting}[language=bash, style=mystyle, caption={Creating the time capsule.}, label=listing:encapsulate]
encapsulator --encapsulate sarah/experiment sarah.R fig1_biplot_v2.png fig2_biplot.png
\end{lstlisting}
Once \texttt{encapsulator} has finished building the time capsule, all that is left to do is to upload it to Sarah's \texttt{Vagrant} cloud account.

\begin{figure}{t}
	\centering
	\includegraphics[width=\columnwidth]{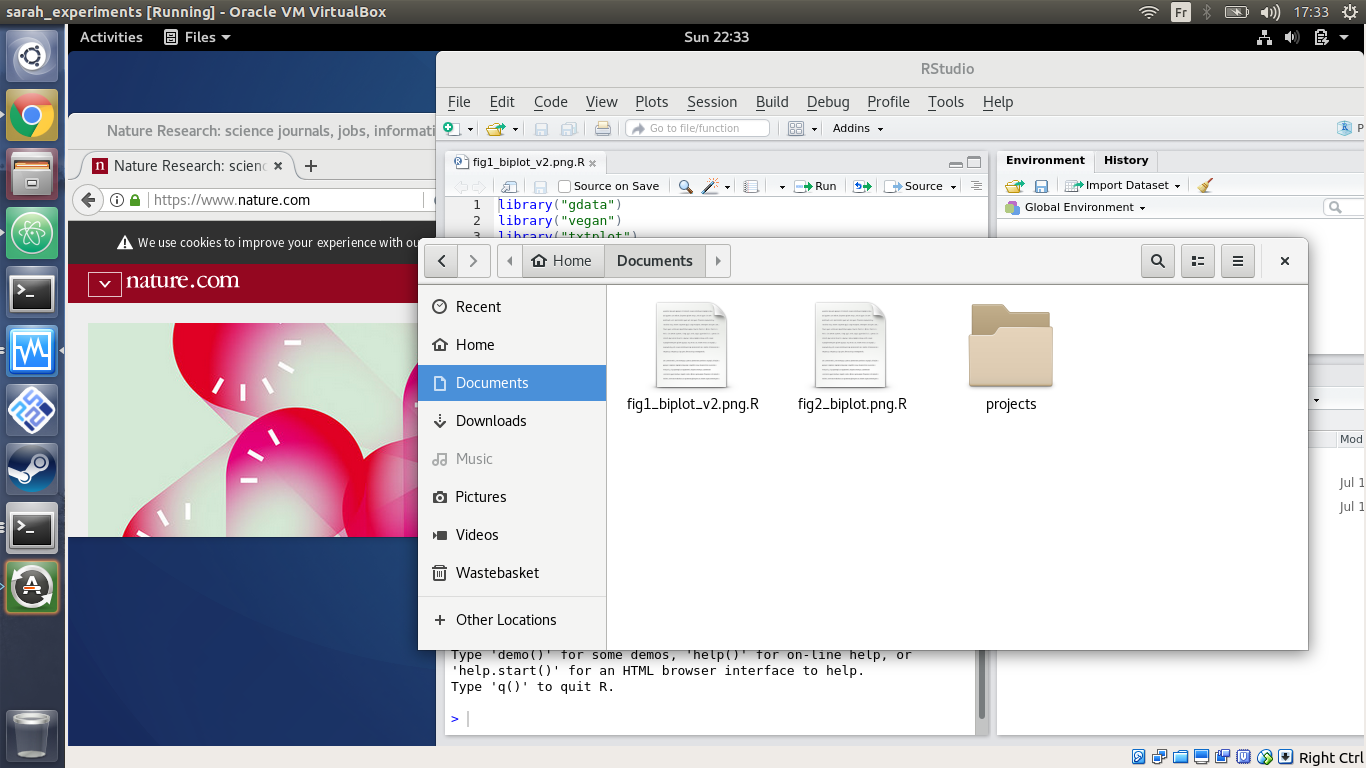}
	\caption{The \emph{time-capsule} running on Prof. O'Brien's machine.}
	\label{img:scenario:vm}
\end{figure}

A few months later, Prof. O'Brien is reviewing Sarah's paper and wants to understand her analysis. He sees that Sarah has used \texttt{encapsulator} to share her work. As Sarah did in her workflow to produce the published results, he can easily install it on a Linux machine:
\begin{lstlisting}[language=bash, style=mystyle, caption={Installing \texttt{encapsulator} and its dependencies}, label=listing:install2]
sudo apt install ruby
gem install encapsulator
encapsulator --install ubuntu
\end{lstlisting}
Once it is installed, he retrieves Sarah's work by running:
\begin{lstlisting}[language=bash, style=mystyle, caption={Decapsulating a shared environment.}, label=listing:decapsulate]
encapsulator --decapsulate sarah/experiment
\end{lstlisting}
\texttt{encapsulator} manages the VM download and startup transparently. 
After a short time, a window appears on Prof. O'Brien's desktop presenting him with the virtual desktop shown in \autoref{img:scenario:vm}. 
In this environment, he has access to familiar tools and can work without difficulty. 
Further, the code that he examines for each figure is about a dozen lines of clean code, not Sarah's original 60 lines of messy code. 
Naturally, \texttt{encapsulator} can handle longer and more complex scripts.

John reads Sarah's article five years after its publication. 
Using the same sequence of commands that Prof. O'Brien used, he is able to get the time capsule running on his laptop, and the environment in the VM is identical to what it was at the time of publication. 
John can get to work easily without worrying about the problem of outdated dependencies (\eg old library versions that are no longer available for download).
 
\section{Challenges}
\label{sec:challenges}
\noindgras{Domain specific environment: }
Our time capsule comes with a generic environment, including some tools generally used for data analysis to provide an easy-to-use, familiar interface. 
In future versions, based on domain-scientist feedback, we will provide platforms containing standard toolsets specific to domains (\eg “ecology”, “genetics”, “chemistry”, \etc).

\noindgras{Time-capsule OS: }
The current version of \texttt{encapsulator} uses Linux, in particular the package management system, to build a time capsule. 
Although a large number of tools used by scientists are available on Windows, Mac, and Linux, some tools may be available only on specific platforms. 
Furthermore, distributing Mac and Windows capsules introduces licensing issues (proprietary software in research is a complex topic~\cite{gambardella2006proprietary}). 
At this stage, one can build a capsule on any platform, but the capsule itself is Linux-based. 
This may not pose a major obstacle for domain scientists whose analytical workflows occur almost entirely within an IDE, such as RStudio, since these IDEs are supported on all major operating systems and appear nearly identical across platforms.

\noindgras{Language support: }
Our current prototype supports only the R programming language. 
We intend to incorporate support for additional languages used in data analysis, including Python and provenance capture libraries such as \texttt{noWorkflow}~\cite{murta2014noworkflow}.
Because \texttt{encapsulator} uses the PROV-JSON standard for data provenance, any provenance capture tool with a statement-level granularity for any language could be used to generate a capsule. 
Furthermore, it should be possible to support individual workflows that use multiple languages, which are becoming more common in some domains.

\noindgras{Integration with IDEs: }
Although they are relatively simple to use, a command-line interfaces are daunting to some users. 
We are investigating integrating \texttt{encapsulator} into existing, commonly-used IDEs, such as an encapsulator add-in for \texttt{RStudio}, a common IDE for \texttt{R} (see \url{https://rstudio.github.io/rstudioaddins/}). 
Many researchers use spreadsheet programs for their data management and analysis. 
Although the feasibility and sufficiency of capturing provenance for such workflows has been demonstrated~\cite{Asuncion2011}, and encapsulation is therefore also theoretically possible, we argue that these methods are inherently unstable since they typically rely on proprietary software with complex underlying data structures. 
Additionally, best practices for data science typically conflict with spreadsheet-based workflows that tend to lead to informal, and often inaccurate, data management and analysis.

\noindgras{Out-of-tree libraries: }
Many obscure libraries may not be available through the package management system, either a specific Linux distribution or a programming language, such as \texttt{CRAN} (see \url{https://cran.r-project.org/}) for \texttt{R} packages. 
We are investigating ways to handle such library dependencies. 
Those that do not have dependencies are relatively easy to handle by building and installing the package during the encapsulation process. 
Others that use alternative package managers, such as Bioconductor (see \url{https://www.bioconductor.org/}), are also relatively easy to handle. 
However, those with complex third-party dependencies without formal definitions are more difficult to support.

\noindgras{Non-deterministic processes: }
Some scripts use pseudo-random-number generators and two runs may not produce identical results. 
We plan to incorporate the ability to reproduce such results in a future release once the provenance capture system records random values; however, a more serious issue is non-determinism introduced by concurrency. 
This could be ameliorated during the curation phase by producing scripts that enforce ordering. 
It might be preferable to enhance how we assess whether a given result produced within the time capsule is correct. 
In the current proof of concept, the results must be identical to those produced on the host machine. 
However, it might be reasonable to verify that the results meet some statistical property instead (\eg within $\delta$ of the original results). 
We recognize that this is not a trivial task and that significant investigation is required to determine a suitable path forward.

\noindgras{Long-term archival: }
There are two major assumptions that \texttt{encapsulator} makes about availability of a time capsule for long-term archival: 
1) the continued existence of the Vagrant cloud; 
and 2) x86-64 virtualization. 
The first issue can be addressed by replicating the time capsule in a trusted archival repository. 
One option that we plan to explore in future work is to publish the time capsule in a \texttt{Dataverse} repository as a ``replication dataset'', assigning automatically a DOI and minimal citation metadata and generating a formal persistent data citation for the time capsule. 
The second issue is more complex, so the answer is speculative. 
Virtualization depends on the remaining life span of the x86-64 architecture and whether the concerned time capsule will have any relevance after that. 
This last point is an interesting issue to ponder, as preservation of our digital world is an issue~\cite{lee2002state} that goes beyond science and reproducibility. 
Artifacts of our modern culture are already disappearing (e.g., video games and digital publications), which is an important socio-cultural issue beyond the scope of our current project.

\noindgras{Container support: }
Although we claim that tools such as Docker are not ideal to reduce the technical barriers to reproducibility for scientists, they are useful for automating the repetition of results. 
As \texttt{Vagrant} supports container provisioning, \texttt{encapsulator} could handle such targets. 
However, one should also remember that while containers are lighter, they are not as self-contained as virtual machines. 
Indeed, containers run over the kernel of their host machine; if change to the kernel were to affect results then reproducibility could not be guaranteed.

\section{Conclusion}
\label{sec:conclusion}
We introduce \texttt{encapsulator}, a sophisticated yet simple toolbox that uses the provenance of computational data analysis to produce a \emph{time capsule} in which computational workflows can be re-run and modified. 
This tool is designed to require minimal overhead for integration into a user's workflow and limited technical expertise. 
When viewed within the context of increasing computational demands of all disciplines, \texttt{encapsulator} provides a key tool for facilitating transparent research at a crucial time for science.
 
\section*{Acknowledgment}
\noindent This work was supported by the US National Science Foundation grant SSI-1450277 \emph{End-to-End Provenance} and grant ACI-1448123 \emph{Citation++}. 
More details about those projects is available at \url{https://projects.iq.harvard.edu/provenance-at-harvard}.

Our reviewers were Prof. Lorena Barba (School of Engineering and Applied Science, George Washington University) and Prof. Carl Boettiger (Department of Environmental Science, Policy and Management, University of California Berkeley). 
They both helped to clarify the terminology used around reproducibility. Prof Boettiger helped us to clarify the extent of the provenance captured.
 
\section*{Software Engineering Practices}
\noindent All software presented in this paper is open-source under GPL v3, and available at \url{http://provtools.org/} or directly through GitHub (\url{https://github.com/ProvTools}). 
The latest version (at the time of submission) can be referenced with the DOI: \emph{10.5281/zenodo.1199232} and is distributed via the RubyGems service (\url{https://rubygems.org/gems/encapsulator}). 
The software presented in this paper remains under development and is subject to change. 
Matthew K. Lau should be contacted for any additional information about the ProvTools ecosystem. 
Further details about continuous integration and engineering practices are available in the README.md files of the individual components. 
\bibliographystyle{IEEEtrans}
\bibliography{biblio}

\newpage

\appendix
\begin{lstlisting}[language=R, style=mystyle, label=apdx:original, caption={Original ``messy'' code.}]
### Messy code is a fabricated example
### intended to capture the essentials
### of a typical, lazy scripter's R code.
### It is, however, tremendously more
### organized than the vast majority of
### scripts.

### Depencies are loaded throughout the script.

### Also, some depencies that are loaded are often
### not used anymore but are still present.
library('gdata')

### Read data from some random file path
### Here, a relative path is being used, but
### typically, file paths are given from root.
data.16 <- read.csv("projects/2016/july_biomass_survey.csv")

### Some datasets are loaded and no longer used.
### Like this one
data.16.2 <- read.csv('projects/data_forestplot/dataset_v2_june_from_collaborator1.csv')

### Create a bunch of intermediate objects
data.v1.1to4 <- data.16[,1:4]
data.v1.1to4. <- data.v1.1to4
data.v1.1to4 <- data.v1.1to4 * 2
data.v1.1to4.2 <- data.v1.1to4 * 2
data.16[,1:4] <- data.v1.1to4.2
library('vegan')
d1 <- vegdist(data.16[,1:2])
d2 <- vegdist(data.16[,2:3])

### Conduct some analyses
mant1 <- mantel(d1,d2)
mant2 <- mantel(d2,d1)
mant11 <- mantel(d1,d1)
fit1 <- lm(Sepal.Length~Sepal.Width,data=data.16)
lm.summary.1 <- summary(fit1)

### write some data to file
write.csv(data.v1.1to4,'projects/data_forestplot/save1.csv',row.names = F)

### write some analyses to file
capture.output(lm.summary.1, file="analysis_forest/anova_table_1.txt")

### write some figures to file
### Here's another random, unused package
library('txtplot')

png('figures_1/fig1_biplot.png')
plot(data.16[,1:2])
dev.off()

png('figures_1/fig1_biplot_t2.png')
plot(data.16[,1:2]*2)
dev.off()

png('figures_2/fig2_biplot.png')
plot(data.16[,2:3])
dev.off()
\end{lstlisting} 
\begin{lstlisting}[language=R, style=mystyle, label=apdx:curated1, caption={Curated code for figure 1.}]
data.16 <- read.csv("projects/2016/july_biomass_survey.csv")
data.v1.1to4 <- data.16[, 1:4]
data.v1.1to4 <- data.v1.1to4 * 2
data.v1.1to4.2 <- data.v1.1to4 * 2
data.16[, 1:4] <- data.v1.1to4.2
png("figures_1/fig1_biplot_v2.png")
plot(data.16[, 1:2] * 2)
dev.off()
\end{lstlisting}

\begin{lstlisting}[language=R, style=mystyle, label=apdx:curated2, caption={Curated code for figure 2.}]
data.16 <- read.csv("projects/2016/july_biomass_survey.csv")
data.v1.1to4 <- data.16[, 1:4]
data.v1.1to4 <- data.v1.1to4 * 2
data.v1.1to4.2 <- data.v1.1to4 * 2
data.16[, 1:4] <- data.v1.1to4.2
png("figures_2/fig2_biplot.png")
plot(data.16[, 2:3])
dev.off()
\end{lstlisting} 
\end{document}